\documentclass[aps,pre,twocolumn,superscriptaddress,showpacs]{revtex4-1}

\usepackage{graphicx}
\usepackage{amsfonts,amssymb}
\usepackage{amsmath}

\newcommand{\Prob}{\mathbb{P}}

\begin{document}


\title{Another phase transition in the Axelrod model} 


\author{Alex Stivala}
\email[]{stivalaa@unimelb.edu.au}
\homepage[]{https://sites.google.com/site/alexdstivala/}
\affiliation{Melbourne School of Psychological Sciences, The University of Melbourne, VIC 3010, Australia}

\author{Paul Keeler}
\email[]{keeler@wias-berlin.de}
\affiliation{Weierstrass Institute, Berlin 10117, Germany}


\date{\today}

\begin{abstract}
Axelrod's model of cultural dissemination, despite its apparent
simplicity, demonstrates complex behavior that has been of much
interest in statistical physics. Despite the many variations and
extensions of the model that have been investigated, a systematic
investigation of the effects of changing the size of the neighborhood
on the lattice in which interactions can occur has not been made. Here
we investigate the effect of varying the radius $R$ of the von Neumann
neighborhood in which agents can interact. We show, in addition to the
well-known phase transition at the critical value of $q$, the number
of traits, another phase transition at a critical value of $R$, and
draw a $q$ -- $R$ phase diagram for the Axelrod model on a square
lattice. In addition, we present a mean-field approximation of the
model in which behavior on an infinite lattice can be analyzed.
\end{abstract}



\maketitle

\section{Introduction}

The Axelrod model of cultural dissemination \citep{axelrod97} is an
apparently simple model of cultural diffusion, in which ``culture'' is
modeled as a discrete vector (of length $F$), a multivariate property
possessed by an agent at each of the $N$ sites on a fully occupied
finite square lattice. Agents interact with their lattice neighbors,
and the dynamics of the model are based on the two principles of
homophily and social influence. The former means that agents prefer to
interact with similar others, while the latter means that agents, when
they interact, become more similar. Despite this apparent simplicity,
in fact the model displays a rich dynamic behavior, and does not
inevitably converge to a state in which all agents have the same
culture. Rather it will converge either to a monocultural state, or a
multicultural state, depending on the model parameters. The Axelrod
model has come to be of great interest in statistical physics, with a
number of variations and analyses conducted. A review from a
statistical physics perspective can be found in \citet{castellano09},
and more recent reviews from different perspectives in
\citet{kashima16chapter,sirbu17}.

One of the best-known features of the Axelrod model is the
nonequilibrium phase transition between the monocultural (ordered) and
multicultural (disordered) states, controlled by the value of $q$, the
number of traits (possible values of each vector element)
\cite{castellano00,gandica13}.  A number of variations and extensions of the
model have been proposed, including an external field (modeling a
``mass media'' effect), noise, and interaction via complex networks
rather than a lattice.

External influence on culture vectors, in the form of a ``generalized
other'' was first introduced by \citet{shibanai01}. Further work on
external influence on culture vectors, or mass media effect, considers
an external field which acts to cause features to become more similar
to the external culture vector with a certain probability
\citep{gonzalez-avella05,gonzalez-avella10,candia08,mazzitello07,rodriguez09,rodriguez10,peres10,gandica13},
or variations such as nonuniform or local fields
\citep{gonzalez-avella06,peres12} or fields with adaptive features
\citep{pinto16}. Counterintuitively, these mass media effects were
found to actually increase cultural diversity rather than result in
further homogenization, an effect explained by local homogenizing
interactions causing the absorbing state to be less fragmented than
when interacting with the external field only, the latter case
actually resulting in more, rather than less, diversity
\citep{peres11}.

The effect of noise, or ``cultural drift'', foreshadowed by
\citet[p. 221]{axelrod97}, in the form of random perturbations of
cultural features, has been examined
\citep{klemm03,parisi03,klemm05,centola07,desanctis09,flache11,gandica13}. A
sufficiently small level of noise actually promotes monoculture,
while too high a level of noise prevents stable cultural regions from
forming (an ``anomic'' state, as described by \citet{centola07}). In fact, there is another phase transition induced by the
noise rate \citep{klemm03}.  Another form of noise, in the form of
random error in determining cultural similarity between agents, has also
been investigated \citep{desanctis09,flache11}. Noise is also
incorporated in various other extensions of the Axelrod model
\citep{candia08,mazzitello07,battiston16,desanctis09,stivala16b,ulloa16}.

Rather than interacting with the neighbors on a lattice, neighborhoods
defined by complex networks have also been investigated, including
both static \citep{klemm03a,guerra10,gandica11,battiston16,reia16b} and
coevolving networks
\citep{centola07,vazquez07,gracia-lazaro11,pfau13}.  The use of
complex networks rather than a lattice results in the phase transition
controlled by the value of $q$ still existing, albeit possibly with a
different critical value. The effect of network topology on
the phase transition driven by noise has also been investigated \citep{kim11}.

Another extension of the Axelrod model is the incorporation of
multilateral influence, that is, interaction between more than
two agents \citep{flache11,rodriguez10}. Multilateral influence allows
diversity to be sustained in the presence of noise, when with dyadic
influence it would collapse to monoculture or anomie \citep{flache11}
--- that is, it removes the phase transition controlled by the noise
rate described by \citet{klemm03}.

Although most investigations of the Axelrod model and its extensions
have been purely through computational experiments, a number of papers
have used either mean-field analysis, or proved rigorous results
mathematically.  The original description of the phase transition
controlled by $q$ used mean-field analysis \citep{castellano00}, as
have some other papers \citep{vilone02,vazquez07b,gandica16}.  A
rigorous mathematical analysis is much more challenging, and has so
far mostly been restricted to the one-dimensional case
\citep{lanchier12a,lanchier12b,lanchier13,lanchier15}, with the
exception of \citet{li14thesis}, who proves results for the usual
two-dimensional model.  The critical behavior of the order parameter
has also been investigated quantitatively for the case of $F=2$ on the
square lattice and small-world networks \citep{reia16b}.
Computational experiments have also been used to investigate the
relationship between the lattice area and the number of cultures
\citep{barbosa09} and thermodynamic quantities such as temperature,
energy, and entropy \citep{villegas-febres08}.  For the
one-dimensional case, \citet{gandica13} propose a thermodynamic
version of the Axelrod model and demonstrate its equivalence to a
coupled Potts model, as well as analyzing its behavior with respect to
noise and an external field. An Axelrod-like model with $F=2$ on a
two-dimensional lattice is analyzed in the asymptotic case
of $N \rightarrow \infty$ by \citet{genzor15}.

Other extensions and variations of the Axelrod model include bounded
confidence and metric features \citep{desanctis09}, agent migration
\citep{gracialazaro09,gracialazaro11a,pfau13,stivala16a}, extended
conservativeness (a preference for the last source of cultural
information) \citep{dybiec12}, surface tension \citep{pace14},
cultural repulsion \citep{radillo-diaz09}, the presence of some agents
with constant culture vectors \citep{reia16,tucci16}, having one or
more features constant on some \citep{singh12} or all
\citep{stivala16a} agents, using empirical \citep{valori12,stivala14}
or simulated \citep{stivala14,babeanu16} rather than uniform random
initial culture vectors, comparing mass media model predictions to
empirical data on a mass media campaign \citep{mazzitello07},
coupling two Axelrod models through global
fields \citep{gonzalez-avella12,gonzalez-avella14}, combining the
Axelrod model with a spatial public goods game \citep{stivala16b},
modeling diffusion of innovations by adding a new trait on a feature
\cite{tilles15}, and even using it as a heuristic for an optimization
problem \citep{fontanari10}.

In addition to the earliest phase diagrams showing just $q$ and the
order parameter \citep{castellano00} or the noise rate $r$ and the
order parameter \citep{klemm03,flache11}, the following phase
diagrams, derived from either simulation experiments, or mean-field
analysis (or both), have been drawn for the Axelrod model and various
extensions (notation may be changed from the original papers for
consistency): $q$ -- $B$ where $B$ is external field strength
\citep{gonzalez-avella05,gonzalez-avella06,gonzalez-avella10,gandica11,mazzitello07};
$r$ -- $\nu$ and $B$ -- $\nu$ where $r$ is noise rate, and $\nu$ is a
parameter controlling the network clustering structure
\citep{candia08}; $q$ -- $o$ where $o$ is the degree of overlap
between the layers of a multilayer network \citep{battiston16};
$\theta$ -- $q$ where $\theta$ is the ``bounded confidence'' threshold
(minimum cultural similarity required for interaction)
\citep{desanctis09}; $F$ -- $q$ for the one-dimensional case
\citep{vilone02}; $\kappa$ -- $q$ where $\kappa$ is the fraction of
``persistent agents'' or ``opinion leaders'' (those with a constant
culture vector) \citep{reia16,tucci16}.

\citet{klemm03a} show a $p$ -- $q$ phase diagram where $p$ is the
rewiring probability on small-world network, and also plot the
relationship between the order parameter (largest region size) and
$k_{\max}/N$ where $k_{\max}$ is maximum node degree in a structured
scale-free network. In the small-world network, the phase transition
still exists and is shifted by the degree of disorder of the network.
In random scale-free networks, the transition disappears in the
thermodynamic limit, but in structured scale-free networks the phase
transition still exists.
\citet{klemm03b} examine the nature of the phase transition in the 
one- and two-dimensional cases, while \citet{hawick13} investigates
in addition three- and four-dimensional systems as well as triangular
and hexagonal lattices.

Despite these extensive investigations into various aspects of the
Axelrod model and its variants, there has been a surprising lack of
systematic investigation of the effect of increasing the neighborhood
size, or ``range of interaction'' \citep{axelrod97} on a simple
Axelrod model with dyadic interaction on a square lattice. This is
despite Axelrod himself discussing the issue briefly
\citep[p.~213]{axelrod97} and conducting experiments with
neighborhoods of size 8 and 12, finding that these result in fewer
stable regions than the original von Neumann neighborhood (size
4). \citet{flache11}, in their model with multilateral influence, use
a larger von Neumann neighborhood size, justifying it as empirically
more plausible and a more conservative test of the preservation of
cultural diversity \citep[pp.~974-975]{flache11}.  Their extended
model makes use of the larger neighborhood as its multilateral social
influence uses more than two agents in an interaction, however all
their experiments, including those reproducing the dyadic
(interpersonal) influence model with noise of \citet{klemm03}, fix the
radius at $R=6$, a precedent followed in a subsequent paper
\citep{ulloa16}, while another model using a larger neighborhood for
multilateral interactions fixes the radius at $R=2$
\citep{stivala16b}.

\citet{vazquez07b} investigate, for the special case $F=2$, the
Axelrod model on a regular random graph using a mean-field analysis,
giving an analytic explanation for the non-monotonic time dependence
of the number of active links. Increasing the coordination number may
be considered to be similar to increasing the neighborhood size on a
lattice with fixed coordination number --- in both cases all agents
have the same number of ``neighbors'' (aside from edge effects in the
case of finite lattices), which increases monotonically with the
coordination number or von Neumann radius
respectively. \citet{vazquez07b} find that larger coordination numbers
give better agreement between their master equation and Axelrod model
simulations, but do not describe a phase transition controlled by the
coordination number.

Here we investigate the effect of varying the radius of the von
Neumann neighborhood in which agents can interact, and find another
phase transition in the Axelrod model at a critical value of the
radius $R$, as well as the well-known phase transition at a critical
value of $q$ \citep{castellano00}, and draw a $q$ -- $R$ phase diagram
for the Axelrod model on a square lattice.

\section{Model}

Each of the $N$ agents on the fully occupied $L \times L$ lattice 
($N = L^2$) has an $F$-dimensional culture vector ($F \geq 2$)
$\sigma_i=(\sigma_{i,1},\dots,\sigma_{i,F})$ for all $1 \leq i \leq N$. Each
entry of the cultural vector represents a feature and takes a single
value from $1$ to $q$, so, more precisely, $\sigma_{i,f} \in
\{1,\dots,q\}$ for all $1 \leq i \leq L^2$ and $1 \leq f \leq F$.
Each of the $F$ elements is referred to as a ``feature'', and $q$ is
known as the number of ``traits''.
 The
cultural similarity of two agents is the number of features they have
in common.  If element $f$ of the culture vector belonging to agent
$i$ is $\sigma_{i,f}$, then the cultural similarity $0 \leq c(i,j) \leq 1$
of two agents $i$ and $j$ is a normalized Hamming similarity
\begin{equation}
c(i,j) = \frac{1}{F}\sum_{k=1}^F \delta_{\sigma_{i,k},\sigma_{j,k}}
\end{equation}
where $\delta_{x,y}$ is the Kronecker delta function.

\begin{figure}
  \includegraphics[width=0.48\textwidth]{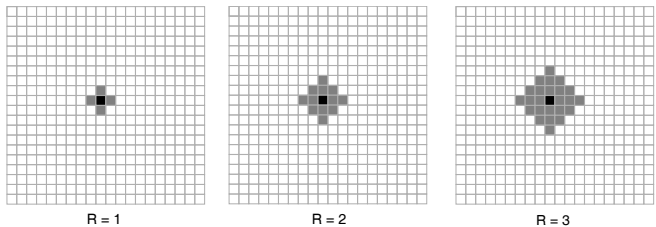}%
  \caption{\label{fig:neighborhoods}Von Neumann neighborhoods of 
    radius $R=1$, $R=2$, and $R=3$. The focal agent is shown in black
    and the von Neumann neighborhood for that agent in gray.}
\end{figure}

An agent can interact with its neighbors, traditionally (as was
originally used by \citet{axelrod97}, for example), defined as the von
Neumann neighborhood, that is, the four (north, south, east, west)
surrounding cells on the lattice, so the number of potentially
interacting agents is the lattice coordination number $g=5$. Here we
extend this to larger von Neumann neighborhoods by increasing the
radius $R$, that is, extending the neighborhood to all cells within a
given Manhattan distance, as was done by \citet{flache11}. This is
illustrated in Figure~\ref{fig:neighborhoods}.  Hence the number of
potentially interacting agents (the focal agent and all its neighbors)
in the von Neumann neighborhood with radius $R$ is now $g(R) = 2R(R+1)+1$
\citep{wolfram,A001844} at most (we do not use periodic boundary
conditions).

Initially, the agents are assigned uniform random culture vectors.
The dynamics of the model are as follows.  A focal agent $a$ is chosen
at random, and another agent $b$ from the radius $R$ von Neumann
neighborhood is also chosen at random.  With probability proportional
to their cultural similarity (the number of features on which they
have identical traits), the two agents $a$ and $b$ interact.  This
interaction results in a randomly chosen feature on $a$ whose value is
different from that on $b$ being changed to $b$'s value. This process
is repeated until an absorbing, or frozen, state is reached. In this
state, no more change is possible, because all agents' neighbors have
either identical or completely distinct (no features in common, so no
interaction can occur) culture vectors.

In the absorbing state, the agents form cultural regions, or clusters.
Within the cluster, all agents have identical culture vectors.  Then
the average size of the largest cluster, $\left<S_{\max}\right>/L^2$
is used as the order parameter \citep{castellano00,klemm03a,castellano09},
separating the ordered and disordered phases. In a monocultural
(ordered) state, $\left<S_{\max}\right>/L^2 \approx 1$, a single
cultural region covers almost the entire lattice; in a multicultural
(disordered) state, multiple cultural regions exist.
Other order parameters that have been used include 
the number of cultural domains \citep{axelrod97,flache11}, mean density of 
cultural domains \citep{peres15}, entropy \citep{radillo-diaz12},
overlap between neighboring sites \citep{klemm03b}, and
activity (number of changes) per agent \citep{reia15}.

Source code for the model (implemented in C++ and Python with MPI 
\cite{dalcin08}) is available from
\url{https://sites.google.com/site/alexdstivala/home/axelrod_qrphase/}.

\section{Results}

\begin{figure}
  \includegraphics[width=0.48\textwidth]{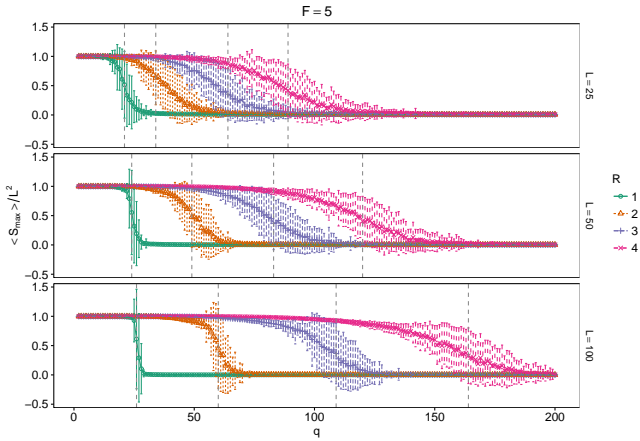}%
  \caption{\label{fig:q} The order parameter
    $\left<S_{\max}\right>/L^2$ (largest region size) plotted against
    the number of traits $q$ for the Axelrod model for $F=5$,
    four different values of the von Neumann radius $R$, and
    three lattice sizes. Each data point
    is the average over 50 independent runs and error bars show the
    95\% confidence interval. Vertical dashed lines show the critical
    value of $q$, where the variance of the order parameter is
    largest.}
\end{figure}

Figure~\ref{fig:q} shows the order parameter (largest region size)
plotted against $q$ for $F=5$, on three different lattice sizes.  It
is apparent that, as the size of the von Neumann neighborhood is
increased, the critical value of $q$ also increases.  That is, by
allowing a larger range of interactions, a larger scope of cultural
possibilities is required in order for a multicultural absorbing
state to exist. Increasing the lattice size has a similar effect, although,
as we shall show in Section~\ref{sec:meanfield}, there is still a finite
critical value of $q$ in the limit of an infinite lattice.

\begin{figure}
  \includegraphics[width=0.48\textwidth]{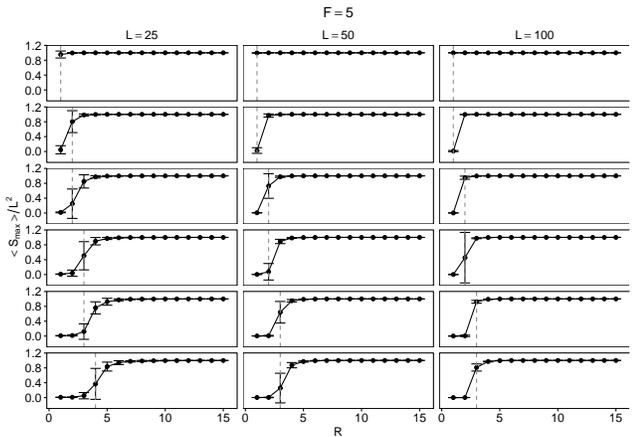}%
  \caption{\label{fig:radius} The order parameter
    $\left<S_{\max}\right>/L^2$ (largest region size) plotted against
    the von Neumann radius $R$ for the Axelrod model for $F=5$,
    some different values of $q$, and
    three lattice sizes. Each data point
    is the average over 50 independent runs and error bars show the
    95\% confidence interval. Vertical dashed lines show the critical
    value of $R$, where the variance of the order parameter is
    largest.}
\end{figure}

Figure~\ref{fig:radius} shows the order parameter (largest region
size) plotted against the von Neumann radius for $F=5$, various values
of $q$, and three different lattice sizes. In each case (apart from
the smallest value of $q$, in which a monocultural state always
prevails), there is a phase transition visible between a multicultural
state (for $R$ less than a critical value) and a monocultural state.
Note that when $R$ is sufficiently large relative to the lattice size
$L$, every agent has every other agent in its von Neumann
neighborhood, and hence the situation is equivalent to a complete
graph or a well-mixed population (or ``soup''
\citep[p.~132]{axtell96}). In this situation, it has long been known
that heterogeneity cannot be sustained \citep{axtell96,axelrod97}.
Fig.~\ref{fig:radius} shows that there appears to be a phase
transition controlled by $R$, between the multicultural phase and the
monocultural phase. As the size of the neighborhood increases, so does
the probability of an agent finding another agent with at least one feature
in common with which to interact, and hence local convergence can
happen in larger neighborhoods, resulting in larger cultural
regions. However this does not result, at the absorbing state (for a
fixed value of $q$), in a gradual increase in maximum cultural region
size from a completely fragmented state to a monocultural state.
Rather, global polarization (a multicultural absorbing state) still
occurs for sufficiently small $R$, but at the critical value of the
radius $R_c$ there is a phase transition so that for neighborhoods
defined by $R > R_c$ a monocultural state prevails.

\begin{figure}
  \includegraphics[width=0.48\textwidth]{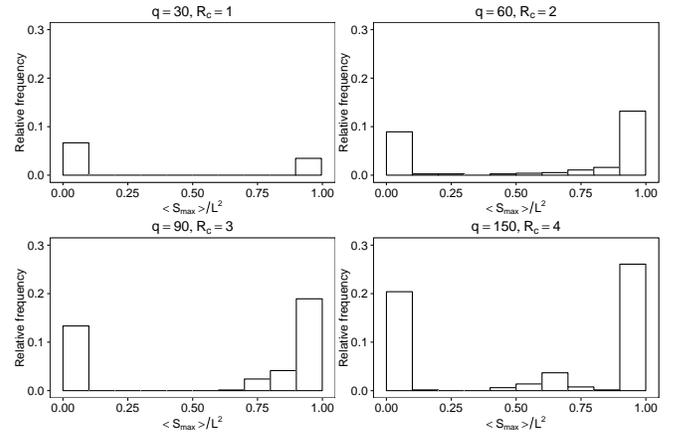}%
    \caption{\label{fig:histogram}Distribution of the order parameter
      at the critical radius for some different values of $q$, with
      $F=5$, $L=100$.  Each distribution is from 50 independent runs.}
\end{figure}

This phase transition is further apparent in
Figure~\ref{fig:histogram}, which shows histograms of the distribution
of the order parameter (largest region size) at the critical radius
for some different values of $q$. That is, for each value of $q$, the
radius $R_c$ at which the variance of the order parameter is greatest.
This shows the bistability of the order parameter at the critical
radius, where the two extreme values are equally probable
\citep{klemm03a}.

\begin{figure}
  \includegraphics[width=0.48\textwidth]{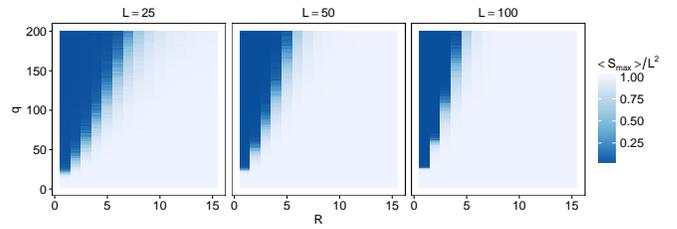}%
  \caption{\label{fig:phase_diagram}$q$ -- $R$ phase diagram showing
    the order parameter $\left<S_{\max}\right>/L^2$ for the Axelrod model for
    $F=5$ and three lattice sizes ($L=25$, $L=50$, and $L=100$). Each data point is colored according to the size of the largest region  $\left<S_{\max}\right>/L^2$ averaged over 50 independent runs. }
\end{figure}

\begin{figure}
  \includegraphics[width=0.48\textwidth]{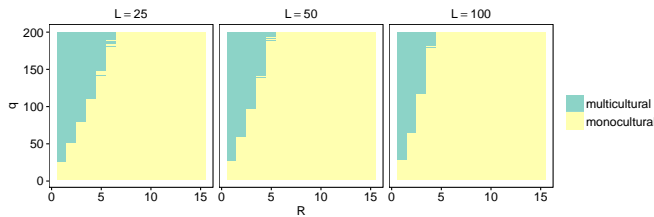}%
  \caption{\label{fig:phase_diagram_2color}$q$ -- $R$ phase diagram
    for the Axelrod model for $F=5$ and three lattice sizes ($L=25$,
    $L=50$, and $L=100$).  As in \citet{klemm03a}, the
    arbitrary, but small, value of 0.1 is used as the value of the
    order parameter to plot the critical value of $q$ separating the
    monocultural and multicultural region for each value of the von
    Neumann neighborhood radius $R$.}
\end{figure}

Figure~\ref{fig:phase_diagram} colors points on the $q$ -- $R$ plane
according to the value of the order parameter, resulting in $q$ -- $R$
phase diagram.  A multicultural state only results for sufficiently
large values of $q$ and small values of $R$.
Figure~\ref{fig:phase_diagram_2color} shows the phase transition more
clearly, with the multicultural states in the upper left of the plane
and the monocultural states in the bottom right.

\section{Mean-field analysis}
\label{sec:meanfield}

We detail the mean-field analysis carried out by \citet{castellano00}
who gave a differential equation. In the mean-field setting, we focus
on the bonds between sites (or agents) located on an infinite lattice,
so we can assume that each site and its von Neumann neighborhood
consists of exactly $g(R)$ sites.  The infinite lattice setting
naturally implies that we do not consider edge effects.

For a single, randomly chosen bond between two sites, we let $P_m(t)$
be the probability that the bond is of type $m$ at time $t$, so both
sites of the bond share $m$ common features, while $F-m$ features are
different. If the randomly chosen bond is connected to sites $i$ and
$j$, then
$$m= \#\{\sigma_{i,f} = \sigma_{j,f}:f\in \{1,\dots, q\}\}.$$

At time $t=0$, we denote by $\rho_0$ the probability of a single
feature of any two sites being common, so $\rho_0=
\Prob(\sigma_{i,f}=\sigma_{j,f})$.  If the
features are distributed uniformly from $1$ to $q$, then
$\rho_0=1/q$. It is sometimes assumed that the features have a Poisson
distribution  \citep{castellano00,vilone02,barbosa09,radillo-diaz12,peres15}
with mean $q$, so then application of the Skellam
distribution gives $\rho_0= e^{-2q}I_0(2q)$, where $I_0$ is a modified
Bessel function of the first kind.  For the single bond, the number of
common features is a binomial random variable, so
\[
P_m(0)= \binom {F} {m} \rho_0^m (1-\rho_0)^{F-m}.
\]


\citet{castellano00} derived a master equation, also known as a
forward equation, given by
\begin{widetext}
\begin{equation}\label{master}
\frac{dP_m(t)}{dt}= \sum_{k=1}^{F-1} \frac{k}{F} P_k(t) \left[\delta_{m,k+1}-\delta_{m,k} +(g-1) \sum_{n=0}^{F} \left( P_n(t)W_{n,m}^{(k)} (t)- P_m(t) W_{m,n}^{(k)} (t) \right )  \right],
\end{equation}
\end{widetext}
where $W_{n,m}^{(k)}(t)$ is the
probability that an $n$-type bond becomes an $m$-type bond due to the
updating of a $k$-type neighbor bond \citep{supplemental_material}.  This equation is only defined
for $1\leq m \leq F$, but naturally the probabilities sum to one,
giving
\[
P_0(t)=1-\sum_{m=1}^{F} P_m(t) .
\]

For $1<m \leq F $, we show \citep{supplemental_material} that  the
master equation or, rather, the set of nonlinear differential
equations~(\ref{master}) can be re-written as
\begin{align}\label{mastersimple}
\frac{dP_m(t)}{dt}= & \left[ \frac{m-1}{F} P_{m-1}(t) -\frac{m}{F} P_m(t) \right] \nonumber \\
&+(g-1)\left[  P_{m-1}(t)W_{m-1,m}^{(k)}(t) \right.  \nonumber \\
&- P_m(t) W_{m,m-1}^{(k)}(t)  \nonumber \\ 
&+  \left.  P_{m+1}(t)W_{m+1,m}^{(k)}(t) \right. \nonumber \\
&- \left. P_m(t) W_{m,m+1}^{(k)}(t) \right ]  \sum_{k=1}^{F-1}  \frac{k}{F} P_k(t)    ,
\end{align}
and zeroth differential equation is
\begin{equation}\label{mastersimple0}
\frac{dP_0(t)}{dt} = -\sum_{m=1}^F \frac{dP_m(t)}{dt} .
\end{equation}

As in \citet{castellano00}, we can investigate model dynamics within
the mean-field treatment by studying the density 
$n_a = \sum_{i=1}^{F-1}P_{i}$ of active bonds, that is a bond across which at
least one feature is different and one the same.  Hence in an
absorbing (frozen) state, $n_a = 0$. In the mean-field analysis, since
an infinite lattice is assumed, $n_a = 0$ only when a multicultural
absorbing state is reached; as noted by \citet{castellano00}, the
coarsening process by which a monocultural state is formed lasts
indefinitely on an infinite lattice.

\begin{figure}
  \includegraphics[width=0.48\textwidth]{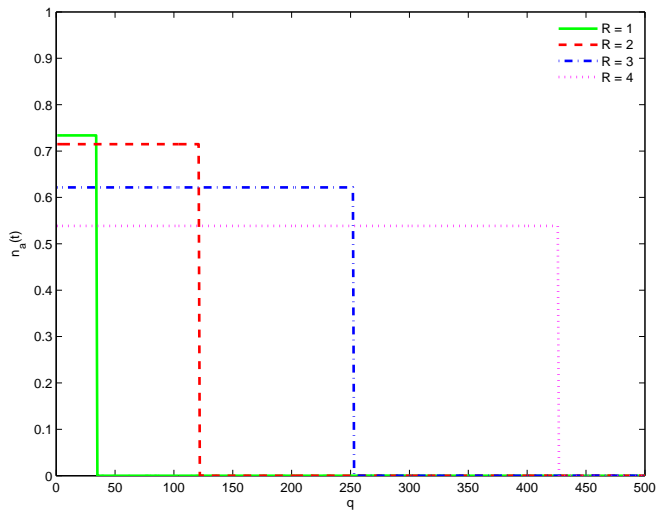}%
  \caption{\label{fig:meanfield_q_radius}Phase diagram within the
    mean-field approximation for some different values of the von
    Neumann radius $R$. The value of $n_a(t)$ (shown at $t=10^3$) is
    obtained by numerical integration
    of~(\ref{mastersimple}) and (\ref{mastersimple0}).}
\end{figure}

Figure~\ref{fig:meanfield_q_radius} plots the number of active bonds
against the value of $q$ for some different values of $R$ within the
mean-field approximation.  It can be seen that the behavior is
qualitatively the same as that shown in Fig.~\ref{fig:q} for the
simulations on finite lattices: the critical value of $q$ is higher
for larger neighborhood sizes. On finite lattices, larger lattice
sizes also increase the critical value of $q$ for a given neighborhood
size, however on an infinite lattice, there is still a finite
critical value of $q$ for a given neighborhood size. This suggests
that, if the lattice size in the simulation could be increased further
(a very computationally demanding process), eventually the critical
values would approach those obtained in the mean-field approximation.

\section{Conclusion}

The original Axelrod model had agents only interact with their
immediate neighbors on a lattice, modeling the assumption of that
geographic proximity largely determines the possibility of
interaction. Subsequent work has extended this to neighbors on complex
networks, or allowed agent migration, or assumed a well-mixed
population (infinite-ranged social interactions) on the assumption
that online interactions are making this assumption more realistic
\citep{valori12}.

Despite these, and other, increasingly sophisticated modifications of
the Axelrod model, however, an examination of the consequences of
simply extending the lattice (von Neumann) neighborhood had not been
carried out.  We have done so, and shown another phase transition in
the model, controlled by the von Neumann radius $R$, as well as the
well-known phase transition at the critical value of $q$, and drawn a
$q$ -- $R$ phase diagram.  We have also used a mean-field analysis to
analyze the behavior on an infinite lattice.

These results show that, as well as the value of $q$, the ``scope of
cultural possibilities'' \citep{axelrod97}, having a critical value
above which a multicultural state prevails, there is also a critical
value of the radius of interaction, above which a monocultural state
prevails.  This simply says that, rather unsurprisingly, a world in
which people can only interact with their immediate neighbors is (for
a fixed value of $q$), more likely to remain multicultural than one in
which people can interact with those further away. Given this
inevitability of a monocultural state for large enough
``neighborhoods'', it might be more useful to consider alternative
measurements of cultural diversity, such as the ``long term cultural
diversity'' measured using the curve plotting the number of final
cultural domains against the initial number of connected cultural
components, as the bounded confidence threshold is varied, as
described by \citet{valori12} (where a well-mixed population was
assumed, and hence a monocultural state results for when the bounded
confidence threshold is zero).  An obvious extension of this work is
to examine the behavior of the Axelrod model on complex networks where
the neighborhood is extended to all agents within paths of length $R$
on the network.

\appendix*

\begin{acknowledgments}
Work by A.S. was supported in part by the Asian Office of Aerospace
Research and Development (AOARD) Grant No.~FA2386-15-1-4020 and the
Australian Research Council (ARC) Grant No.~DP130100845.
P.K. acknowledges the support of Leibniz program ``Probabilistic
Methods for Mobile AdHoc Networks'' and ARC Centre of Excellence for
the Mathematical and Statistical Frontiers (ACEMS) Grant
No.~CE140100049,
and thanks Prof.~Peter G. Taylor for helpful discussion
and for the invitation to visit Melbourne. This research was supported
by Victorian Life Sciences Computation Initiative (VLSCI) grant number
VR0261 on its Peak Computing Facility at the University of Melbourne,
an initiative of the Victorian Government, Australia.  We also used
the University of Melbourne ITS Research Services high performance
computing facility and support services.
\end{acknowledgments}

%

\end{document}